\documentstyle[11pt,fleqn]{article}
\begin{document}

\begin{titlepage}

\begin{flushright}
Freiburg--THEP 95/25\\
November 1995
\end{flushright}
\vspace{1.5cm}

\begin{center}
\large\bf
{\LARGE\bf The Higgs resonance shape in gluon fusion:
           Heavy Higgs effects}\\[1cm]
\rm
{A. Ghinculov}\\
{and}\\
{J.J. van der Bij}\\[.5cm]

{\em Albert--Ludwigs--Universit\"{a}t Freiburg,
           Fakult\"{a}t f\"{u}r Physik}\\
      {\em Hermann--Herder Str.3, D-79104 Freiburg, Germany}\\[1.5cm]

\end{center}
\normalsize

\begin{abstract}
We study the influence of two--loop radiative corrections
of enhanced electroweak
strength on Higgs production at the LHC.
We consider Higgs production by the gluon
fusion mechanism, with the subsequent decay of the Higgs boson into
a pair of Z bosons, and incorporate the resonance shape corrections up to
order $(g^2 \, m_H^2 / m_W^2)^2$. We take into account the full
$g g \rightarrow Z Z$ process and the $q \bar{q} \rightarrow Z Z$
background, as well as the subsequent decay of the Z pair into
leptons. We also discuss the theoretical uncertainty related to
the use of the equivalence theorem in this process.
\end{abstract}

\vspace{3cm}

\end{titlepage}


\section{Introduction}

The arguably most important task of the Large Hadron Collider is
to provide evidence regarding the nature of the mechanism responsible
for the spontaneous breaking of the electroweak symmetry. This is
crucial for understanding the physics at the electroweak scale and may
give hints about the dynamics underlying the
standard model at higher energies.

The simplest model of spontaneous symmetry breaking, the Higgs sector
of the minimal standard model, will be testable at the LHC up to an energy
scale of $\sim 1$ TeV. This covers the mass range where the Higgs
particle is believed to make sense as a fundamental field.
This is related to the triviality
of the Higgs sector, which is a nonperturbative problem.
Lattice simulations
of the $\phi^4$ theory seem to indicate an upper bound
of the order of 650 GeV for $m_H$, beyond which the Higgs mass would
be larger than the cutoff scale \cite{hasenfratz}.
Therefore a strongly interacting Higgs sector
would not exist. However, these results must be interpreted with
caution. There are indications that they depend on the  lattice
regularization, and in fact the triviality bound may be considerably higher
\cite{heller}.
Also the presence of a large Yukawa coupling may play
a r\^ole because it relaxes the bound derived from the
position of the Landau pole.
Unfortunately, it is difficult to check this nonperturbatively
because of notorious problems with treating fermions on the lattice
in the presence of Yukawa interactions \cite{stephenson}.
One cannot exclude the possibility
of a Higgs mass of the order of 1 TeV.

The Higgs boson is produced in hadron collisions via the gluon
fusion and the vector boson fusion processes.
With $m_t \approx 180$ GeV, the gluon fusion process dominates
unless the Higgs is very massive, of the order of 1 TeV \cite{kunszt}.
Heavy Higgs bosons mainly
decay into pairs of vector bosons. The decay channel
$H \rightarrow Z Z \rightarrow 2 (l^+ l^-)$
provides a clean signal if $m_H$ is smaller than $\sim 800$ GeV.
For a heavier Higgs, the channel
$H \rightarrow Z Z \rightarrow l^+ l^- \nu \bar{\nu}$
must be included. The Higgs then appears as a Jacobian peak
in the $p_T$ distribution of the $Z$ bosons.
To approach the 1 TeV mass range,
one needs to consider the
$H \rightarrow Z Z \rightarrow l^+ l^- + 2 \, jets$
and
$H \rightarrow W^+ W^-$
channels, which are affected by large backgrounds.

The Higgs production mechanisms at hadron colliders
were studied extensively at
leading order \cite{kunszt}.
The NLO QCD corrections to the gluon fusion
process were calculated recently, and indicate a substantial increase
of the Higgs production \cite {spira1,spira2}. It is the purpose of
this paper to study the NNLO corrections of enhanced
electroweak strength to Higgs production via gluon fusion,
with the Higgs subsequently decaying into a Z pair.

The large $m_H$ range is affected by potentially large electroweak
radiative corrections due to the strong selfinteraction of the Higgs
sector. For instance, the radiative corrections
of increased electroweak strength
to the Higgs decay into vector bosons
are known at two--loop order \cite{2loop:htoww}, and
are of the order of 24\% for $m_H \sim 900$ GeV. It is therefore
legitimate to ask what is the effect of such corrections on the
production and decay of heavy Higgs bosons.

A consequence of the Slavnov--Taylor identities of the theory is
the equivalence theorem. This theorem relates in the high energy limit
the amplitudes
with external longitudinal vector bosons $V_L$ and the corresponding
amplitudes with the $V_L$ replaced by Goldstone bosons
\cite{cornwall}--\cite{he:ET}.
The equivalence theorem is a useful tool for calculating leading contributions
in the Higgs mass in a simple way, but within this context the resonance
region should be treated carefully. The equivalence theorem holds order
by order in perturbation theory, while in the resonance region a Dyson
summation is needed which contains incomplete contributions
from higher orders. If one fails to carry out the calculation consistently
at the given order in the coupling constant, the result will differ
from the corresponding leading $m_H$ contributions
calculated in the full electroweak
theory (see for instance ref. \cite{yuan} and \cite{valwill2}, and references
therein).

Related to the large coupling domain is also the problem of
unitarity violation when the Higgs mass exceeds certain limits.
Unitarity breaking effects are always present within perturbation
theory and are formally of higher order in the coupling constant.
Numerically, they may become important if the coupling constant is large,
and raise questions about the relevance of calculations
of physical processes.

A few approaches were proposed in the literature to deal with such problems.
To mention only two of them, ref. \cite{valwill1} proposes the use of
a momentum dependent width of the Higgs
in the resonant propagator instead of a constant width in order to reduce
the size of the unitarity violations in longitudinal vector boson
scattering. Ref. \cite{seymour}
notes that this procedure reproduces the correct result in the resonance
region but has a bad behavior at higher energies, where the nonresonant
propagator should be used. Therefore it proposes
a summation of the full tree level Goldstone boson
scattering amplitude rather than a Dyson summation of only the Higgs
selfenergy in order to interpolate between the two regions.

These approaches amount in fact to changes
in the incomplete contributions of higher order in the coupling constant.
One should see them as leading order calculations.
They are not complete at one--loop order, and their use is largely
motivated by the fact that they lead to a solution which has the desired
numerical behaviour in the specific process considered.
For instance,
using the momentum dependent width in the Higgs resonant propagator gives
a numerically well behaved amplitude in ref. \cite{valwill1},
but would lead to
an unwanted result in the case of fermion--fermion scattering.
Indeed, the one--loop momentum dependent selfenergy used in
ref. \cite{valwill1} would result in a shift in the position
of the Higgs resonance in fermion scattering at leading order
in $m_H$, which in fact is
absent in the full one--loop radiative corrections \cite{2loop:method}.
In this case the constant width is obviously the better choice.

A systematic
treatment would instead require a complete calculation of radiative
corrections to the given process. This ensures both that the
equivalence theorem is valid up to the desired order, and that
the unitarity violating contributions are of a higher order
in the coupling constant compared to the tree amplitude, which is the
best one can do within perturbation theory.

This paper is devoted to the study of the two--loop
radiative corrections of enhanced electroweak strength
to the $gg \rightarrow ZZ$ process.
By using the equivalence theorem, we derive in the
following section the two--loop corrections to the shape of the Higgs
resonance in the related process
$t \bar{t} \rightarrow H \rightarrow z z$.
We then study the influence of these corrections on Higgs searches
by means of a Monte--Carlo simulation of the full
$g g \rightarrow Z Z \rightarrow 4 f$
process and of the
$q \bar{q} \rightarrow Z Z \rightarrow 4 f$
background at LHC.


\section{Resonance shape corrections}

In this section we derive the NNLO radiative corrections to the
scattering process
$t \bar{t} \rightarrow H \rightarrow zz$, where $z$ are Goldstone bosons.
We are interested only in radiative corrections of
enhanced electroweak strength, that is, corrections which grow like
$m_H^2$ at one--loop order and like $m_H^4$ at two--loop.

In the following, we denote by $\lambda$ the selfcoupling of the Higgs field,

\begin{equation}
  \lambda = \frac{g^2}{16 \pi^2} \frac{m_H^2}{m_W^2}  \nonumber
              \; \; ,
\end{equation}
with $g^{2} = 4 \sqrt{2} \, m_{W}^{2} \, G_{F}$,
$G_{F} = 1.16637 \cdot 10^{-5} \; GeV^{-2}$, and $m_{W} = 80.22 \; GeV$.

The radiative corrections of leading order in $m_H$ to the
$t \bar{t} \rightarrow H \rightarrow zz$
process are depicted in fig. 1. They consist of the
radiative correction to the Yukawa coupling $V_{Ht\bar{t}}$,
which is momentum independent \cite{2loop:htoferm,kniehl},
the momentum dependent correction to the
$Hzz$ vertex $V_{Hzz}(s)$, and the correction to the Higgs propagator
$\Pi (s)$.
In the resonance region one needs to perform an all--order Dyson
summation of the proper selfenergy. Far from the peak, this resonant
propagator will differ from the nonresummed two--loop selfenergy
insertion by contributions of order $\lambda^3$.

At leading order, the diagram of fig. 1 reads simply

\begin{equation}
  \frac{1}{s - m_H^2 + i m_H \Gamma_H}
              \; \; ,
\end{equation}
up to the tree level $H t \bar{t}$ and $Hzz$ couplings,
which are constants independent of $s$.
$\Gamma_H$ in eq. 2 is the tree level Higgs width.
This expression is correct at leading order,
{\em i.e.}, ${\cal O}(\lambda^{-1})$
on the resonance and
${\cal O}(1)$
far from the resonance.

The radiative corrections change this expression into:

\begin{equation}
   V_{Ht\bar{t}} \,
   \frac{1}{s - m_H^2 + i m_H \Pi (s)}  \,
   V_{Hzz} (s)
              \; \; .
\end{equation}

We are primarily interested in radiative corrections in the
resonance region. We therefore perform a momentum expansion around
$s = m_H^2$:

\begin{eqnarray}
\Pi (s) & = & \Pi (m_H^2) + \frac{s-m_H^2}{m_H^2}     \Pi^{\prime}
                          + \frac{(s-m_H^2)^2}{m_H^4} \Pi^{\prime \prime}
			  + \dots
               \nonumber \\
V_{Hzz}(s) & = & V_{Hzz}(m_H^2) + \frac{s-m_H^2}{m_H^2} V_{Hzz}^{\prime}
               + \frac{(s-m_H^2)^2}{m_H^4} V_{Hzz}^{\prime \prime}
	       + \dots
      \; \; ,
\end{eqnarray}

For treating the radiative corrections consistently as an expansion in the
coupling constant $\lambda$, one notes that in the resonance region
the quantity $s-m_H^2$ is of the order of $m_H \Gamma_H$, that is,
${\cal O}(\lambda)$. This must be taken into account together
with the loop expansion of the quantities
$V_{Ht\bar{t}}$,
$\Pi (m_H^2)$,
$V_{Hzz}(m_H^2)$,
$\Pi^{\prime}$,
$V_{Hzz}^{\prime}$, \dots .
To have the complete NNLO amplitude, one needs to keep the following
contributions:

\begin{eqnarray}
V_{Ht\bar{t}} & = & 1 + V_{Ht\bar{t}}^{(1-loop)}
                      + V_{Ht\bar{t}}^{(2-loop)}
	              + {\cal O}(\lambda^3)
                \nonumber \\
V_{Hzz}(s)    & = & 1 + V_{Hzz}^{(1-loop)}(m_H^2)
                      + V_{Hzz}^{(2-loop)}(m_H^2)
              \nonumber \\
              & &     + \frac{s-m_H^2}{m_H^2} V_{Hzz}^{\prime \, (1-loop)}
	              + {\cal O}(\lambda^3)
                \nonumber \\
\Pi (s)       & = &     \Pi^{(1-loop)} (m_H^2)
                      + \Pi^{(2-loop)} (m_H^2)
                      + \Pi^{(3-loop)} (m_H^2)
                \nonumber \\
              & &     + \frac{s-m_H^2}{m_H^2}     \Pi^{\prime \, (2-loop)}
		      + {\cal O}(\lambda^4)
      \; \; .
\end{eqnarray}

The correction to the Yukawa coupling reads \cite{2loop:htoferm}
\cite{kniehl}:

\begin{eqnarray}
V_{Ht\bar{t}}^{(1-loop)} & = &
    ( \frac{13}{16} - \frac{\pi \, \sqrt{3}}{8} ) \, \lambda
  = .132325 \, \lambda
                \nonumber \\
V_{Ht\bar{t}}^{(2-loop)} & = & (-.26387 \pm 1.3 \cdot 10^{-4}) \, \lambda^2
      \; \; .
\end{eqnarray}

The correction to the $Hzz$ vertex was calculated in ref. \cite{2loop:htoww}:

\begin{eqnarray}
V_{Hzz}^{(1-loop)} (m_H^2) & = &
     \left[ \frac{19}{16} + \frac{5 \, \pi^2}{48}
          - \frac{3 \, \sqrt{3} \, \pi}{8}
     + i \, \pi \left( \frac{\log{2}}{4} - \frac{5}{8} \right) \right]
     \, \lambda
                \nonumber \\
 & = &  \left( .17505951 - i \, 1.4190989 \right) \, \lambda
                \nonumber \\
V_{Hzz}^{(2-loop)} (m_H^2) & = & -
       \left[   (.53673 \pm 4.1 \cdot 10^{-4} )
         + i \, (.32811 \pm 3.1 \cdot 10^{-4} )  \right]
       \, \lambda^2
      \; \; .
\end{eqnarray}

By evaluating the one--loop diagrams which contribute to the $H \rightarrow zz$
decay, on finds also:

\begin{eqnarray}
V_{Hzz}^{\prime \, (1-loop)} & = &
     \left[ 1 + \frac{ \sqrt{3} \, \pi}{12} - \frac{5 \, \pi^2}{48}
     + i \, \pi \left( \frac{1}{8} - \frac{\log{2}}{4}  \right) \right]
     \, \lambda
                \nonumber \\
 & = &  \left( .42536605  - i \, .15169744 \right) \, \lambda
      \; \; .
\end{eqnarray}

Let us now turn to the correction to the Higgs propagator.

At the order in which we are working, the quantity $\Pi (s)$ is
the imaginary part of the proper Higgs selfenergy. The real part
of the Higgs proper selfenergy at $s=m_H^2$ and its first derivative
with respect to $s$ are absorbed into the Higgs mass counterterm and
wave function renormalization, respectively.
The real part of the selfenergy may
contribute an imaginary piece to eq. 5 only at ${\cal O}(\lambda^4)$,
through the quantity $\Pi^{\prime \prime \, (2-loop)}$.
Therefore $\Pi(m_H^2)$ is the total width of the Higgs including the
appropriate radiative corrections.

At tree level, the main decay channels of a heavy Higgs are given by:

\begin{eqnarray}
\Gamma^{(tree)}_{H \rightarrow W^+ W^-} & = &
 \frac{g^2}{64 \pi} \frac{m_H^3}{m_W^2}
 \left[ 1 - 4 \frac{m_W^2}{m_H^2} \right]^{1/2}  \times
  \nonumber \\  & &
 \left[ 1 - 4 \frac{m_W^2}{m_H^2} + 12 \frac{m_W^4}{m_H^4} \right]
    \nonumber \\
\Gamma^{(tree)}_{H \rightarrow Z^0 Z^0} & = &
 \frac{g^2}{128 \pi} \frac{m_H^3}{m_W^2}
 \left[ 1 - 4 \frac{m_Z^2}{m_H^2} \right]^{1/2}  \times
   \nonumber \\  & &
 \left[ 1 - 4 \frac{m_Z^2}{m_H^2} + 12 \frac{m_Z^4}{m_H^4} \right]
    \nonumber \\
\Gamma^{(tree)}_{H \rightarrow t \bar{t}} & = &
 \frac{3 g^2}{32 \pi} \frac{m_H \, m_t^2}{m_W^2}
 \left[ 1 - 4 \frac{m_t^2}{m_H^2} \right]^{3/2}
      \; \; .
\end{eqnarray}

The radiative corrections to the decay widths of eqns. 9 are known at
two--loop level in the heavy Higgs approximation
\cite{2loop:htoferm,kniehl,2loop:htoww}:

\begin{eqnarray}
\lefteqn{ \Gamma_{H \rightarrow W^+ W^- \, , \, Z^0 Z^0} \, =  \,
          \Gamma^{(tree)}_{H \rightarrow W^+ W^- \, , \, Z^0 Z^0} \, \times }
 \nonumber \\
 & &  \left[
 1 + \lambda
     \left( \frac{19}{8} + \frac{5 \, \pi^2}{24}
          - \frac{3 \, \sqrt{3} \, \pi}{4}
    \right)
   + \lambda^2
     \left( \, .97103 \pm 8.2 \cdot 10^{-4} \, \right)
 \right]
    \nonumber \\
 & = & \Gamma^{(tree)}_{H \rightarrow W^+ W^- \, , \, Z^0 Z^0} \,
 \left[
 1 + .350119 \, \lambda
   + \left( \, .97103 \pm 8.2 \cdot 10^{-4} \, \right) \lambda^2
 \right]
    \nonumber \\
\lefteqn{ \Gamma_{H \rightarrow t \bar{t}} \, =  \,
                      \Gamma^{(tree)}_{H \rightarrow t \bar{t}} \, \times
     \left[
 1 + \lambda
     \left(  \frac{13}{8} - \frac{\pi \, \sqrt{3}}{4} \right)
   - \lambda^2
     \left( \, .51023 \pm 2.5 \cdot 10^{-4} \, \right)
 \right]
 }
 \nonumber \\
 & = & \Gamma^{(tree)}_{H \rightarrow t \bar{t}} \,
  \left[
 1 + .264650 \, \lambda
   - \left( \, .51023 \pm 2.5 \cdot 10^{-4} \, \right) \lambda^2
 \right]
      \; \; .
\end{eqnarray}

Note that in eqns. 10 some incomplete subleading contributions are present in
the radiative corrections, as discussed in ref. \cite{2loop:htoww}. They appear
if one multiplies the full tree level width by the radiative correction factor
calculated in the leading $m_H$ approximation. These terms are of the same
order in the coupling constant as the theoretical uncertainty related
to the use of the equivalence theorem while calculating radiative corrections.
It is thus not possible to decide unambiguously whether it is better to keep
them or to drop them without calculating the complete subleading contributions
explicitly. Numerically, this ambiguity is at 1\% level at most.
As such it can be safely neglected.

With these results, one can identify in eq. 5:

\begin{eqnarray}
\lefteqn{  \Pi^{(1-loop)} (m_H^2)
         + \Pi^{(2-loop)} (m_H^2)
         + \Pi^{(3-loop)} (m_H^2)  \, =}  \nonumber \\
& &   \Gamma_{H \rightarrow W^+ W^-}
    + \Gamma_{H \rightarrow Z^0 Z^0}
    + \Gamma_{H \rightarrow t \bar{t}}
    + \Gamma_{H \rightarrow 4w \, , \, wwzz \, , \, 4z}
    \; \; ,
\end{eqnarray}
with $\Gamma_{H \rightarrow W^+ W^-}$, $\Gamma_{H \rightarrow Z^0 Z^0}$
and $\Gamma_{H \rightarrow t\bar{t}}$
given by eqns. 10.

The term $\Gamma_{H \rightarrow 4w \, , \, wwzz \, , \, 4z}$
originates from the
three--loop cut diagrams shown in fig. 2. All other cut diagrams of
three--loop selfenergy diagrams are allready contained in
$\Gamma^{(2-loop)}_{H \rightarrow W^+ W^- \, , \, Z^0 Z^0}$.

Including or dropping $\Gamma_{H \rightarrow t\bar{t}}$ in eq. 11
is in principle irrelevant at the order in which we are working
because this is a contribution of higher order in the top quark mass.
In practice this is not a large effect because the $t\bar{t}$ branching
ratio is small and because of the partial cancellation between the one--loop
and the two--loop corrections to this partial width.

The quantity $\Pi^{\prime \, (2-loop)}$, which also contributes to
eq. 5, was calculated in ref. \cite{2loop:method}:

\begin{eqnarray}
\Pi^{\prime \, (2-loop)} \, = \, m_H \, \lambda^2 \,
  \frac{3 \, \pi}{4} \,
  \left(
  1 + \frac{\pi \, \sqrt{3}}{12} - \frac{5 \, \pi^{2}}{48}
  \right) & = & \nonumber \\
  = \; \;
  1.002245142 \, m_H \, \lambda^2
     \; \; .
\end{eqnarray}

To evaluate $\Gamma_{H \rightarrow 4w \, , \, wwzz \, , \, 4z}$,
let us introduce the following notations:

\begin{eqnarray}
 \Gamma_{H \rightarrow 4z} & = &
                              \varphi_{4z} \, \lambda^2 \, \Gamma_0
    \; \; \; \; \; \; ,
    \nonumber \\
 \Gamma_{H \rightarrow  2z2w }  & = &
                              \varphi_{2z2w} \, \lambda^2 \, \Gamma_0
    \; \; \;  ,
    \nonumber \\
 \Gamma_{H \rightarrow 4w} & = &
                              \varphi_{4w} \, \lambda^2 \, \Gamma_0
    \; \; \; \; \; \; , \; \; \; \; \; \;
 \Gamma_0  =  \frac{3 \, \pi}{8} \, m_H \lambda
    \; \; .
\end{eqnarray}

We calculate the dimensionless factors $\varphi$ by means of a Monte--Carlo
integration over the four--body phase space of the cut diagrams of fig. 2.

The diagrams themselves were calculated through the equivalence theorem,
therefore only the leading $m_H$ terms are kept
at the level of the amplitude. However, when integrating over the phase space,
we allow for a finite mass of the vector bosons by keeping the momenta of the
outgoing Goldstone bosons at an invariant mass equal to the vector
boson masses, $m_Z$ and $m_W$.
This procedure is similar to the way we multiplied in eqns. 10
the full tree level width by the loop corrections obtained via
the equivalence theorem. In both cases the result contains
incomplete subleading terms from the phase space integration
(eq. 10 contains also contributions from the
transversal degrees of freedom of the vector bosons).
This is motivated by the assumption that the phase space factor
is numerically the main subleading effect. To check the reliability of
the equivalence theorem calculation of this process,
we calculated the decay $H \rightarrow 4Z^0$ in the electroweak theory.
We calculated the decay $H \rightarrow 4Z^0_L$ in
the approximation that the longitudinal polarization vectors of
the $Z$ are given by $\epsilon_L(p_\mu)\approx p_\mu /m_Z$, as well
as the complete $H \rightarrow 4Z^0$ decay.

\begin{table}
\begin{tabular}{||c||c|c|c||c|c||}    \hline\hline
$m_{H}[GeV]$ & $\varphi_{4w}^{(ET)}$ & $\varphi_{2z2w}^{(ET)}$
             & $\varphi_{4z}^{(ET)}$ & $\varphi_{4z}^{(Z^0_L)}$
             & $\varphi_{4z}^{(Z^0)}$               \\ \hline\hline
600      & $4.14 \cdot 10^{-5}$ & $1.88 \cdot 10^{-5}$ & $8.97 \cdot 10^{-6}$ &
$9.07 \cdot 10^{-6}$ & $1.69 \cdot 10^{-6}$ \\ \hline
700      & $6.42 \cdot 10^{-5}$ & $3.34 \cdot 10^{-5}$ & $1.58 \cdot 10^{-5}$ &
$1.67 \cdot 10^{-5}$ & $4.67 \cdot 10^{-6}$ \\ \hline
800      & $8.47 \cdot 10^{-5}$ & $4.81 \cdot 10^{-5}$ & $2.22 \cdot 10^{-5}$ &
$2.38 \cdot 10^{-5}$ & $8.74 \cdot 10^{-6}$ \\ \hline
900      & $1.02 \cdot 10^{-4}$ & $6.24 \cdot 10^{-5}$ & $2.79 \cdot 10^{-5}$ &
$3.01 \cdot 10^{-5}$ & $1.33 \cdot 10^{-5}$ \\ \hline
1000     & $1.18 \cdot 10^{-4}$ & $7.53 \cdot 10^{-5}$ & $3.25 \cdot 10^{-5}$ &
$3.51 \cdot 10^{-5}$ & $1.78 \cdot 10^{-5}$ \\ \hline
1100     & $1.31 \cdot 10^{-4}$ & $8.70 \cdot 10^{-5}$ & $3.66 \cdot 10^{-5}$ &
$3.93 \cdot 10^{-5}$ & $2.22 \cdot 10^{-5}$ \\ \hline
1200     & $1.41 \cdot 10^{-4}$ & $9.65 \cdot 10^{-5}$ & $4.01 \cdot 10^{-5}$ &
$4.29 \cdot 10^{-5}$ & $2.62 \cdot 10^{-5}$ \\ \hline
1500     & $1.64 \cdot 10^{-4}$ & $1.20 \cdot 10^{-4}$ & $4.73 \cdot 10^{-5}$ &
$5.00 \cdot 10^{-5}$ & $3.59 \cdot 10^{-5}$ \\ \hline
2000     & $1.87 \cdot 10^{-4}$ & $1.45 \cdot 10^{-4}$ & $5.41 \cdot 10^{-5}$ &
$5.61 \cdot 10^{-5}$ & $4.61 \cdot 10^{-5}$ \\ \hline
3000     & $2.05 \cdot 10^{-4}$ & $1.66 \cdot 10^{-4}$ & $5.93 \cdot 10^{-5}$ &
$6.04 \cdot 10^{-5}$ & $5.51 \cdot 10^{-5}$ \\ \hline
5000     & $2.16 \cdot 10^{-4}$ & $1.80 \cdot 10^{-4}$ & $6.24 \cdot 10^{-5}$ &
$6.29 \cdot 10^{-5}$ & $6.08 \cdot 10^{-5}$ \\ \hline
$\infty$ & $2.23 \cdot 10^{-4}$ & $1.89 \cdot 10^{-4}$ &  $6.41 \cdot 10^{-5}$
& $6.41 \cdot 10^{-5}$ & $6.41 \cdot 10^{-5}$ \\ \hline\hline
\end{tabular}
\caption{The value of the factors $\varphi$ calculated via the equivalence
theorem
($\varphi^{(ET)}$) and in the electroweak theory,
as a function of the Higgs mass.
For $\varphi^{(ET)}$, the momenta of the outgoing Goldstone bosons
have invariant masses $m_Z$ and $m_W$. $\varphi_{4z}^{(Z^0_L)}$
corresponds to the decay $H \rightarrow 4Z^0_L$ calculated in
the electroweak theory in the approximation $\epsilon_L(p_\mu)\approx p_\mu
/m_Z$.
$\varphi_{4z}^{(Z^0)}$ is the exact result, summed over the polarization
states of the $Z$ bosons.
}
\end{table}

The results are given in table 1. The limit $m_W=m_Z=0$, in which the
equivalence theorem is exact, corresponds to $m_H \rightarrow \infty$.
One can see that the equivalence theorem with the phase-space factors
included approximates well
the electroweak decay width in the approximation
$\epsilon_L(p_\mu)\approx p_\mu /m_Z$.
However, the equivalence theorem and the exact $H \rightarrow 4Z^0$
result start to agree only for $m_H$ of the order of 2.5---3 TeV.
The equivalence theorem
is not a good
approximation for $m_H \sim 1$ TeV because $m_H$ is not large enough
compared to the four--particle threshold at $\sim 360$ GeV.

This is to
be contrasted to the situation encountered in one-- and two--loop calculations.
At two--loop level there are only cut diagrams with two massles particles
on the cut lines because the Goldstone bosons are always produced in pairs.
In this case the equivalence theorem works better than at three--loop level,
where diagrams with four massles particles on the cut lines are allowed,
and the threshold in the full electroweak theory is higher.

Generally, for given masses of the vector bosons and for fixed external
momenta, one expects the equivalence theorem to become a progressively
bad approximation as the number of loops increases. The subleading
contributions become increasingly important because of the presence of
multiparticle cuts.

In this calculation, the uncertainty related to
subleading contributions in the
four Goldstone decay
is numerically negligible.
Compared to the two Goldstone channel,
the four Goldstone decay is strongly
suppressed by phase space.

We have now all ingredients needed to evaluate eq. 3 at NNLO.
At lowest order, the expression which we obtain is equivalent to eq. 2.
At the same time, it incorporates the full radiative corrections of
${\cal O}(\lambda)$ and ${\cal O}(\lambda^2)$.
It also contains some incomplete subleading contributions.

In fig. 3 we compare the LO expression to
the NNLO calculation.

Although the purpose of this calculation was primarily
to obtain the radiative corrections to the
shape of the resonance in a consistent way, one notes that the
amplitude is well behaved far from the resonance as well.

Compared to the corrections to the Higgs width, the resonance
shape corrections of fig. 3 are relatively small. This is because
of the partial compensation of the effect of the radiative
corrections to $\Pi(s)$ and to $V_{Hzz}(s)$. For the same reason,
the shift of the resonance peak towards lower momentum is smaller
than the shift in fermionic scattering \cite{2loop:method}.

Finally, let us notice that the NNLO is the lowest order where
nontrivial corrections to the resonance shape occur.
At NLO there is only a correction
to the Higgs width because the momentum dependence of the quantities
$\Pi$ and $V_{Hzz}$ is of higher order in $\lambda$.


\section{Gluon fusion at hadron colliders}

The diagrams which contribute to the process $gg \rightarrow ZZ$
are shown in fig. 4. Among these, only the Higgs production
diagram receives radiative corrections of enhanced electroweak strength.

The leading $m_H$ radiative corrections to the Higgs production
diagram of fig. 4 a) are identical to the corrections to
the process $t \bar{t} \rightarrow H \rightarrow zz$,
which were discussed in the previous section, by the equivalence theorem.
This is true in the limit of large top mass as well.
In this limit the top quark does not decouple, and the $gg \rightarrow H$
subgraph results in the effective interaction
${\cal L}^{eff}_{ggH} = g \, \alpha_s /(24 \, \pi \, m_W) \,
                        H \, G^a_{\mu \nu} \, G^{a \, \mu \nu}$,
which receives the same radiative corrections at leading order in
$m_H$ as the Yukawa coupling.

We incorporated the radiative corrections of the previous section in
a Monte--Carlo event generator to calculate the $gg \rightarrow ZZ$
process and the $q\bar{q} \rightarrow ZZ$ background at LHC.
The subsequent decay of the $Z$ pair into fermions is included
in the narrow width approximation by using the density matrix formalism.
Details of the calculation can be found in ref.
\cite{matsuuravdbij,glovervdbij}.

We take $\sqrt{s}=14.5$ TeV for the CM energy of the LHC, and $m_t=180$ GeV
for the mass of the top quark. We use the MHRS parton distribution
functions \cite{MHRS}, with $\Lambda = 190$ MeV, evolved to the scale
$Q^2=\hat{s}/4$, and $\alpha_S= 12\pi/(23 \log{Q^2/\Lambda^2})$
corresponding to five flavours.
As a rough simulation of the detector geometry, we impose a rapidity cut
on the outgoing leptons of $|y_l|<2.5$, and request that the leptons have
a transverse momentum larger than 20 GeV. The cross sections shown
in the following correspond to four muons in the final state.
The branching ratio should be multiplied by
a factor 4 for muons or electrons in
the final state, and 24 for $l^+l^-\nu\bar{\nu}$
(with these cuts, the cross sections for neutrinos in the final state
will be underestimated to some extent).
One can consider an integrated luminosity of $10^2 fb^{-1}$ to interpret
the results.

The invariant mass spectrum of the $Z$ pair is shown in fig. 5 for different
values of the Higgs mass. The background refers to $m_H \rightarrow \infty$.
The transverse momentum distribution of the $Z$ bosons is shown in fig. 6.
One notices that the way the radiative corrections influence the shape
of the Higgs resonance differs from the effect shown in fig. 3 for the
process $t\bar{t} \rightarrow H \rightarrow zz$. In particular, the
cross section on top of the resonance is slightly increased.
This is due to interference effects with the nonresonant diagrams in
fig. 4.


\section{Conclusions}

We studied the effects of the radiative corrections of enhanced electroweak
strength on Higgs production by gluon fusion at hadron colliders.
The full correction to the Higgs resonance shape at
next--to--next--to--leading order was derived. This was then incorporated
in a Monte--Carlo simulation of the processes $gg \rightarrow ZZ$
and $q \bar{q} \rightarrow ZZ$ at LHC.

Compared to the similar radiative corrections to the $H \rightarrow V_L V_L$
decay, the Higgs resonance shape corrections turn out to be
relatively small. This is
due to cancellations between different contributions. Other than for
the $t \bar{t} \rightarrow H \rightarrow zz$ scattering, in the gluon fusion
process $gg \rightarrow H \rightarrow Z Z$ the
radiative corrections result in an increase of 10---20\% of the
Higgs signal
due to interference effects with the nonresonant diagrams.

We did not address the question of the other
Higgs production mechanism, the vector boson fusion.
The cross section of this process becomes more important as the
mass of the Higgs boson increases. For the vector boson scattering process,
one--loop results are available \cite{veltmanyndurain}
in the large $m_H$ limit.
Results at two--loop order in the large momentum limit
also exist \cite{riesselmann:wscatt}.
A complete two--loop analysis would be difficult to perform
even via the equivalence theorem
because one needs to evaluate two--loop box diagrams at finite
external momenta. This problem deserves further investigation.


\vspace{.5cm}

{\bf Acknowledgement}

This work was supported by the Deutsche Forschungsgemeinschaft (DFG).


\newpage


{\bf Figure captions }

\vspace{2cm}

{\em Fig.1}    The radiative corrections of enhanced electroweak
               strength to the process
	       $t\bar{t} \rightarrow H \rightarrow zz$.

\vspace{.5cm}

{\em Fig.2}    Cut diagrams which contribute to the imaginary part
               of the Higgs selfenergy at three--loop order. These
	       contributions are not included in the two--loop
	       $H \rightarrow zz,w^+w^-$ decay width.

\vspace{.5cm}

{\em Fig.3}    Radiative corrections to the Higgs resonance shape
               in the process $t\bar{t} \rightarrow H \rightarrow zz$
	       for $m_H=850$ GeV. The solid line
	       is the squared absolute value
	       of the leading order (eq. 2). The thin line
	       corresponds to the NNLO expression (eq. 3).

\vspace{.5cm}

{\em Fig.4}    Feynman diagrams for $Z^0$ pair production by gluon fusion.

\vspace{.5cm}

{\em Fig.5}    Invariant mass distribution of the $Z^0$ pairs at LHC.
               The processes considered are
	       $gg \rightarrow ZZ \rightarrow 2(\mu^+\mu^-)$
	       and
	       $q\bar{q} \rightarrow ZZ \rightarrow 2(\mu^+\mu^-)$.
	       We take $\sqrt{s}=14.5$ TeV, and for the outgoing
	       muons we request $p_T>20$ GeV and $|y_l|<2.5$.
               The solid line is the NNLO cross section, the dashed
	       line is the tree level cross section, and the dotted line
	       is the background (no Higgs production diagram).
	       a) shows the total cross section, and b) shows the
	       Higgs signal, with the background subtracted.

\vspace{.5cm}

{\em Fig.6}    Transverse momentum distribution of the $Z^0$ bosons at LHC.
               The processes considered are
	       $gg \rightarrow ZZ \rightarrow 2(\mu^+\mu^-)$
	       and
	       $q\bar{q} \rightarrow ZZ \rightarrow 2(\mu^+\mu^-)$.
	       We take $\sqrt{s}=14.5$ TeV, and for the outgoing
	       muons we request $p_T>20$ GeV and $|y_l|<2.5$.
               The solid line is the NNLO cross section, the dashed
	       line is the tree level cross section, and the dotted line
	       is the background (no Higgs production diagram).
	       a) shows the total cross section, and b) shows the
	       Higgs signal, with the background subtracted.

\end{document}